\documentclass[11pt]{article}

\newcommand{\be}{\begin{equation}}
\newcommand{\ee}{\end{equation}}
\newcommand{\ba}{\begin{eqnarray}}
\newcommand{\ea}{\end{eqnarray}}
\def\Gammabol{{\stackrel{\circ}{\Gamma}}{}}
\def\Abol{{\stackrel{~\circ}{A}}{}}
\def\Bbol{{\stackrel{~\circ}{B}}{}}
\def\Rbol{{\stackrel{\circ}{R}}{}}
\def\Obol{{\stackrel{\circ}{\Omega}}{}}
\def\Gammabol{{\stackrel{\circ}{\Gamma}}{}}
\def\Tbol{{\stackrel{\circ}{\mathcal T}}{}}
\def\Dbol{{\stackrel{\circ}{\mathcal D}}{}}
\def\nabol{{\stackrel{\circ}{\nabla}}{}}

\def\onehalf{{\textstyle{\frac{1}{2}}}}
\def\ihalf{{\textstyle{\frac{i}{2}}}}

\begin{document}
\renewcommand{\thefootnote}{\fnsymbol{footnote}}
\noindent
{\Large \bf General Relativity as a Genuine Connection Theory}\footnote{Dedicated to Jos\'e
Pl\'\i nio Baptista on the occasion of his seventieth birthday.}
\vskip 0.7cm
\noindent
{\bf R.~Aldrovandi, H.~I.~Arcos\footnote{Permanent address: Universidad Tecnol\'ogica
de Pereira, A.A. 97, La Julita, Pereira, Colombia.} and J.~G.~Pereira} \\
{\it Instituto de F\'{\i}sica Te\'orica} \\
{\it Universidade Estadual Paulista} \\
{\it Rua Pamplona 145} \\
 {\it 01405-900 S\~ao Paulo, Brazil}


\vskip 0.8cm
\begin{abstract}
\noindent
The Palatini formulation is used to develop a genuine connection theory for general relativity,
in which the gravitational field is represented by a Lorentz-valued spin connection. The
existence of a tetrad field, given by the Fock--Ivanenko covariant derivative of the
tangent-space  coordinates, implies a coupling between the spin connection and the coordinate
vector-field, which turns out to be the responsible for the onset of curvature. This
connection-coordinate coupling can thus be considered as the very foundation of the
gravitational interaction. The peculiar form of the tetrad field is shown to reduce both
Bianchi identities of general relativity to a single one, which brings this theory closer to
the gauge theories describing the other fundamental interactions of Nature. Some further
properties of this approach are also examined.
\end{abstract}


\section{Introduction}
\setcounter{footnote}{0}
\renewcommand{\thefootnote}{\arabic{footnote}}

Differently from gauge theories~\cite{iz} the fundamental field of general relativity is not a
connection, but a metric tensor---or equivalently, a tetrad field. The difficulties in
quantization have created a continuous interest  in  developing a connection-based formulation
of the theory~\cite{ash1},  which would bring it closer to the gauge theories describing the
other fundamental interactions of Nature and to their successful quantization
techniques~\cite{po}. The standard way of bringing connections to the forefront is to use
Palatini's variational method, by which the connection and the tetrad are independent
variables. The field equations are, accordingly, obtained from independent variations with
respect to both the tetrad and the spin connection. The first variation yields Einstein's
equation, whereas the second yields a constraint equation whose solution defines the
connection in terms of the tetrad.   Once written in terms of the tetrad, however, the
connection loses its  role as fundamental field, and one is led back to the usual metric-based
formulation. In a genuine connection theory the spin connection should keep the role of
fundamental field and
 should not be written in terms of the tetrad. It is the tetrad, as a derived field,  that must
be written in terms of the connection. The last point is the basic difference between  the
construction we are going to consider here and all existing approaches to the so called
connection-based theories of gravity~\cite{ash}.

The reformulation of general relativity as a Lorentz-valued connection theory for gravitation
requires a change in the traditional {\em kinematic paradigm} of the theory. More specifically,
instead of spacetime diffeomorphisms, the fundamental transformations behind a connection-based
formulation of general relativity must be assumed to be the {\em local} Lorentz
group~\cite{cp1}. In order to implement such a change it is necessary to devote special
attention to the Minkowski tangent space ${\mathcal M}$, which is naturally attached to each
point of spacetime ${\mathcal R}$ and on which the Lorentz transformations will take place. In
this way, general relativity can be reinterpreted as a theory emerging from the requirement of
covariance under local Lorentz transformations~\cite{uk}, and accordingly the spin connection
turns up as the basic field representing gravitation. It is important to notice that, as the
Einstein-Hilbert Lagrangian stands on, this reinterpretation entails no change in {\em
dynamics}, but only in the  underlying {\em kinematics} of the theory. Conceptual changes,
however, do show up which lead to fundamental differences with respect to the ordinary metric
formulation. The basic purpose of this work is to study these differences.

\section{Modified Palatini Formulation}

\subsection{The Spin Connection}

We use the Greek alphabet ($\mu$, $\nu$, $\rho, \dots=1,2,3,4$) to denote indices related to
spacetime, and the Latin alphabet ($A,B,C, \dots = 1,2,3,4$) to denote indices related to each
one of the Minkowski tangent spaces, whose metric tensor is chosen to be $\eta_{AB} = {\rm
diag} (+1, -1, -1, -1)$. Using this notation, the spin connection $A_\mu$, a field assuming
values in the Lie algebra of the Lorentz group, is written as
\be %
A_\mu = \onehalf A{}^{AB}{}_\mu \, J_{AB},
\ee %
where $J_{AB}$ are the Lie algebra generators written in some appropriate representation.
Since $A{}^{AB}{}_\mu = - A{}^{BA}{}_\mu$, it automatically preserves the Minkowski
metric:\footnote{The presence of nonmetricity would spoil the anti-symmetry in the first two
indices, and consequently the connection would not be Lorentz-valued.}
\[
\partial_\mu \eta_{AB} - A^C{}_{A \mu} \, \eta_{CB} - A^C{}_{B \mu} \, \eta_{AC} = 0.
\]
The curvature of the connection $A^A{}_{B \mu}$ is
\be
\Omega^A{}_{B \mu \nu} = \partial_\mu {A}^A{}_{B \nu} - \partial_\nu
{A}^A{}_{B \mu} + {A}^A{}_{E \mu} \, {A}^E{}_{B \nu} -
{A}^A{}_{E \nu} \, {A}^E{}_{B \mu}.
\label{cur}
\ee
Denoting by $h_\mu=h^A{}_\mu \partial_A$ a general tetrad field, the torsion of $A^A{}_{B
\mu}$ is written in the form
\be
{\mathcal T}^A{}_{\mu \nu} = \partial_\mu {h}^A{}_{\nu} - \partial_\nu
{h}^A{}_{\mu} + {A}^A{}_{E \mu} \, {h}^E{}_{\nu} -
{A}^A{}_{E \nu} \, {h}^E{}_{\mu}.
\label{tor}
\ee
It is important to remark that curvature and torsion are properties of a
connection~\cite{KoNo}. Notice, however, that in the case of non-soldered bundles, as for
example in Yang-Mills theories, no tetrad exists, and consequently torsion cannot be even
defined. This is completely different from general relativity, whose spin connection, the so
called Ricci coefficient of rotation, has vanishing torsion. Although vanishing, therefore,
torsion is always present in general relativity.

Using the tetrad, a spin connection $A^A{}_{B \mu}$ can be related with the corresponding
spacetime connection $\Gamma^{\rho}{}_{\nu \mu}$ through
\be
\Gamma^{\rho}{}_{\nu \mu} = h_{A}{}^{\rho} \partial_{\mu} h^{A}{}_{\nu} +
h_{A}{}^{\rho} A^{A}{}_{B \mu} h^{B}{}_{\nu}.
\label{geco}
\ee
The inverse relation is, consequently,
\be
A^{A}{}_{B \mu} =
h^{A}{}_{\nu} \partial_{\mu}  h_{B}{}^{\nu} +
h^{A}{}_{\nu} \Gamma^{\nu}{}_{\rho \mu} h_{B}{}^{\rho}.
\label{gsc}
\ee
Equations (\ref{geco}) and (\ref{gsc}) are simply different ways of expressing the property
that the total---that is, acting on both indices---derivative of the tetrad vanishes
identically:
\be
\partial_{\mu} h^{A}{}_{\nu} - \Gamma^{\rho}{}_{\nu \mu} h^{A}{}_{\rho} +
A^{A}{}_{B \mu} h^{B}{}_{\nu} = 0.
\label{todete}
\ee
In what follows, we will denote the magnitudes related with general relativity with an over
``$\circ$''. For example, the Ricci coefficient of rotation will be denoted by $\Abol{}^C{}_{A
\nu}$, whereas its curvature will be $\Obol^A{}_{B \mu \nu}$.

\subsection{Ordinary Palatini Formulation}

The basic gravitational variables in the Palatini framework is the pair ($h_\mu, A_{\mu})$ of
{\it independent} 1-form fields. As is well known, the tetrad field provides an isomorphism
between the tangent space ${\mathcal M} = T_x{\mathcal R}$ at each $x^\mu$ and the algebra of
the translation group, which is also a vector space equipped with the metric $\eta_{AB}$. It
establishes, therefore, a relation between $\eta_{AB}$ and the spacetime metric $g_{\mu \nu}$:
\be
g_{\mu \nu} = \eta_{AB} \, h^A{}_\mu \, h^B{}_\nu.
\label{gmn}
\ee

In the specific case of general relativity, the action of the gravitational field in the
Palatini formulation can be written in the form
\be
S_P =  \frac{1}{4 k^2} \int_{\mathcal R}
{\epsilon}^{\mu\nu\rho\sigma} \epsilon_{ABCD} \,
h^A{}_\mu \, h^B{}_\nu \, \Obol^{CD}{}_{\rho \sigma},
\label{palaction}
\ee
where $k^2=8 \pi G/c^4$, $\epsilon_{ABCD}$ is the totally anti-symmetric Levi--Civita tensor
on ${\mathcal M}$ compatible with $\eta_{AB}$, and
\be
\epsilon_{\mu\nu\rho\sigma} = h^A{}_\mu \, h^B{}_\nu h^C{}_\rho \, h^D{}_\sigma \,
\epsilon_{ABCD},
\ee
with $\epsilon_{0123} = h \equiv \det(h^A{}_\mu)$. In contrast to the usual
Einstein--Hilbert action,
$S_P$ depends on two independent variables. Variation of
$S_P$ with respect to the (inverse) tetrad $h_A{}^\mu$ yields Einstein's equation
\be
\Obol^A{}_\mu - \textstyle{\frac{1}{2}} \; h^A{}_\mu \, \Obol = 0,
\label{einteineq} 
\ee
whereas variation  with respect to the spin connection $\Abol^A{}_{B \mu}$ yields
\be
\partial_\mu h^{A}{}_{\nu} - \partial_\nu
h^{A}{}_{\mu} + [\Abol_\mu, h_\nu]^{A} = 0,
\label{vincu}
\ee
which is a constraint equation determining the vanishing of torsion. It is then usually assumed
that Eq.~(\ref{vincu}) can be solved for $\Abol^A{}_{B \mu}$, in which case it becomes
completely determined by the tetrad: $\Abol_\mu = \Abol_\mu(h_\nu)$. A further restriction to
histories in which the connection is so determined reduces $S_P$ to the ordinary
Einstein--Hilbert action of general relativity,
\be
S_{EH} \equiv S_P(h_\nu, \Abol_\mu(h_\nu)) = \frac{1}{2 k^2} \int_{\mathcal R}
d^4x \; h \; \Rbol,
\label{ehaction}
\ee
where $\Rbol$ is the scalar curvature of $\Abol_\mu = \Abol_\mu(h_\nu)$.

\subsection{Modifying the Palatini Formulation}

The above procedure is not consistent with a genuine connection-based theory for gravitation,
since in such a theory the connection is to be considered as a fundamental, and not a derived
field like $\Abol_\mu(h_\nu)$. If we really want to obtain a formulation for gravity which is
closer to a Yang--Mills theory, the spin connection is to be considered as a fundamental field,
and accordingly the field equation (\ref{vincu}) must be solved for the coframe $h_\nu$, and
not for the connection $\Abol_\mu$. In other words, the coframe is to be completely determined
by the connection: $h_\nu = h_\nu(\Abol_\mu)$. Of course, as far as the theory is kept metric
compatible, and torsion is assumed to vanish, the resulting theory is the same as general
relativity, though written in a different set of field--coordinates. In this case, however, a
restriction to histories on which the coframe is determined in terms of the spin connection
$\Abol_\mu$ does not reduce $S_P$ to the Einstein--Hilbert action. It leads, actually, to a
modified version of the Palatini action, which we indicate as
\be
{S'}_{P} \equiv S_P(h_\nu(\Abol_\mu), \Abol_\mu).
\ee
Although presenting the same dynamics, the resulting theory will have different
features in relation to general relativity. For example, the spacetime connection
\be
\Gammabol^\rho{}_{\mu \nu} = 
h_A{}^\rho \partial_\nu h^A{}_\mu + h_A{}^\rho \Abol^A{}_{B \nu} \, h^B{}_\mu,
\label{lcc}
\ee
despite presenting zero torsion, will never, in this formulation, be written in terms of the
metric or the tetrad. The crucial point of this modified Palatini formulation is then to solve
the constraint equation (\ref{vincu}) for $h^A{}_\mu$ in terms of the spin connection
$\Abol^A{}_{B \mu}$. As we are going to see next, the requirement of local Lorentz covariance
naturally yields such a solution.

\section{Lorentz Transformations}

Let us review some basic properties of the Lorentz transformations. Denoting the Cartesian
Minkowski coordinates by $\{x^A\}$, the most general form of the generators of infinitesimal
Lorentz transformations is~\cite{ramond}
\begin{equation}
J_{AB} = L_{AB} + S_{AB},
\label{fullrep}
\end{equation}
where
\begin{equation}
L_{AB} = i (x_A \partial_B - x_B \partial_A)
\label{loregen}
\end{equation}
is the {\em orbital} part of the generators, and $S_{a b}$ is the {\em spin} part of the
generators, whose explicit form depends on the spin of the representation. The generators
$J_{AB}$ satisfy the commutation relation
\begin{equation}
[J_{AB}, J_{CD}] = i \left(\eta_{BC} \, J_{AD} - \eta_{AC} \, J_{BD} -
\eta_{BD} \, J_{AC} + \eta_{AD} \, J_{BC} \right),
\label{commu}
\end{equation}
which is to be identified with the Lie algebra of the Lorentz group. Each set of
generators $L_{AB}$ and $S_{AB}$ satisfies the same commutation relation as $J_{AB}$,
and these sets commute with each other.

A local---that is, position dependent---infinitesimal Lorentz transformation of Minkowski
space coordinates is usually written with the orbital generators,
\begin{equation} %
\delta_L x^A = - \ihalf \, \epsilon^{CD} \, L_{CD} \, x^A \equiv
- \epsilon^{A}{}_D \, x^D, 
\label{lore}
\end{equation} %
where $\epsilon^{CD} \equiv \epsilon^{CD}(x^\mu)$ is the parameter of the transformation. Now,
due to the transitivity of Minkowski spacetime under translations, every two points related by
a Lorentz transformation can be also related by a translation (though the converse is not
true). In fact, by using the explicit form of $L_{CD}$, the transformation (\ref{lore}) can be
rewritten as 
\be %
\delta_L x^A = - i \, \xi^C \, P_C \, x^A, 
\label{relore}
\ee %
which is a translation with
\be %
\xi^C = \epsilon^C{}_D \, x^D 
\label{constr}
\ee %
as parameter, and $P_C = - i \, \partial_C$ as generator. This means essentially that an
infinitesimal Lorentz transformation of the Minkowski coordinates is formally equivalent to a
translation with $\xi^C$ as parameters.

On the other hand, because the coordinates $x^A$  behave collectively as a vector under Lorentz
transformations, we can interpret their set $\{x^A(x^\mu)\}$ as a vector field. In this case,
the Lorentz generators are those of the (spin) vector representation~\cite{ramond},
\be %
\left( S_{CD}\right)^A{}_B = i \left( \delta_C{}^A \, \eta_{DB} -
\delta_D{}^A \, \eta_{CB} \right),
\label{spinge}
\ee %
which yields
\be %
\delta_S x^A = - \ihalf \, \epsilon^{CD} \, \left(
S_{CD}\right)^A{}_B \, x^B \equiv \epsilon^A{}_D \, x^D.
\label{slore}
\ee %
Therefore, we see that a Lorentz transformation of the Minkowski coordinates written
with the complete generators $J_{CD}$ vanishes identically:
\be %
\delta_J x^A \equiv - \ihalf \, \epsilon^{CD} \, J_{CD} \, x^A = 0.
\label{lore3}
\ee %
Of course, the generators are defined up to a sign. However, provided $L_{ab}$ and
$S_{ab}$ are chosen in such a way to satisfy the same commutation relation, they yield opposite
Lorentz transformations, and consequently a vanishing {\it total} Lorentz transformation. This
result---quite consistent by itself---comes from the fact that, concomitant with the Lorentz
transformation (\ref{slore}) in its vector indices, a vector field $V^A(x)$, for example,
necessarily undergoes the Lorentz transformation~(\ref{lore}) in its arguments, yielding a
fixed point transformation:
\be
\delta V^A(x) \equiv V^{\prime A}(x) - V^A(x) =
- \ihalf \, \epsilon^{CD} \, J_{CD} \, V^A(x).
\ee
In the case of the coordinate itself, which is a Lorentz vector field, both transformations
cancel each other, yielding a vanishing net result. For all other fields, $J_{AB}$
generates a Lorentz transformation at a fixed spacetime point, or equivalently, the change
in the functional form of the field.

\section{Lorentz Covariant Derivative and Coupling Prescription}

Let us consider now a general matter field $\Psi(x^\mu)$, which is a function of the
spacetime coordinates $\{x^\mu\}$. When considering the gravitational interaction, like in
any gauge theory, the relevant transformation is that associated with the change in the
functional form of the field. These transformations, as is well known, are generated by
$J_{AB}$~\cite{ramond}:
\be %
\delta_J \Psi \equiv \Psi^\prime(x) - \Psi(x) = -\ \ihalf \,
\epsilon^{AB} J_{AB} \Psi(x).
\label{lt1}
\ee %
The explicit form of $L_{AB}$ is the same for all fields, whereas that of $S_{AB}$ depends on
the Lorentz representation $\Psi$ belongs to. Notice that the orbital generators $L_{AB}$ are
able to act on the spacetime argument of $\Psi(x^\mu)$, due to the relations
\[
\partial_A = (\partial_A x^\mu) \, \partial_\mu \quad \mbox{and} \quad
\partial_\mu = (\partial_\mu x^A) \, \partial_A.
\]
By using the explicit form of $L_{AB}$, the Lorentz transformation (\ref{lt1}) can be
rewritten in the form
\be %
\delta_J \Psi =
- i \xi^C P_C \Psi - \ihalf \, \epsilon^{AB} S_{AB} \Psi,
\label{lt3}
\ee %
with $\xi^C$ given by Eq.~(\ref{constr}). Again, we see that the {\em orbital} part of the
transformation reduces to a translation, and consequently the Lorentz transformation of a
general field $\Psi$ can be rewritten as a ``translation'' plus a pure spin transformation.
Despite presenting such a particular form, it is important to notice that, because $[P_C,
S_{AB}]=0$, it is not a Poincar\'e, but a genuine Lorentz transformation.

In a connection-based approach to general relativity, the fundamental field representing
gravitation is the spin connection $\Abol_\mu$. Accordingly, the Lorentz covariant derivative
of the matter field $\Psi$, that is, the derivative which is covariant under the Lorentz
transformations generated by the $J_{AB}$'s is~\cite{livro}
\be %
\Dbol{}_C \Psi = \partial_C \Psi + \onehalf \,
\Abol^{AB}{}_C \, \frac{\delta_J \Psi}{\delta \epsilon^{AB}},
\ee %
where $\Abol^{AB}{}_C = \Abol^{AB}{}_\mu \, h_C{}^\mu$. Substituting the transformation
(\ref{lt1}), it becomes
\be %
\Dbol{}_C \Psi =
\partial_C \Psi - \ihalf \, \Abol{}^{AB}{}_C \, J_{AB} \Psi.
\label{locode0}
\ee %
Using the identity
\be
\ihalf \, {\Abol^{AB}}_\mu \, J_{AB} = \ihalf \, {\Abol^{AB}}_\mu \, S_{AB} +
\Bbol^A{}_\mu \, P_A,
\label{nonholo}
\ee
where $\Bbol^A{}_\mu \equiv {\Abol^A}_{B \mu} \, x^B$, the covariant derivative
(\ref{locode0}) can be rewritten in the form
\be %
\Dbol_C \Psi =
h_C{}^\mu \, \Dbol_\mu \Psi,
\label{locode}
\ee %
where
\be %
\Dbol_\mu = \partial_\mu -
\ihalf \, \Abol^{AB}{}_\mu \, S_{AB}
\label{fi}
\ee %
is the usual Fock--Ivanenko covariant derivative operator~\cite{fi}, and $h_C{}^\mu$ is the
inverse of the tetrad~\cite{cp1}
\be %
h^C{}_\mu = \partial_\mu x^C + \Abol{}^C{}_{D \mu} \, x^D.
\label{tetrada}
\ee %
Our initial contention, that the tetrad field should be a derived quantity written in terms
of the spin connection $A{}^C{}_{D \mu}$, is in this way vindicated.

On account of the above results, the coupling of a general matter field to gravitation can
then be accomplished by replacing all ordinary by covariant derivatives:
\be
\partial_C \rightarrow \Dbol_C \equiv
h_C{}^\mu \, \Dbol_\mu.
\ee
The gravitational coupling prescription, therefore, is composed of two parts. The
Fock--Ivanenko derivative accounts for the coupling of the spin of the matter field to
gravitation. This coupling is not universal as it depends on the spin content of the matter
field. On the other hand, the tetrad, which appears in the coupling prescription multiplying
the Fock-Ivanenko derivative, accounts for the coupling of the energy and momentum of the
matter field to gravitation. This part of the coupling prescription is universal in the sense
that all fields in Nature will respond equally to its action. As the nontrivial part of the
tetrad comes from the orbital Lorentz generators, we can say that these generators are the
responsible for the universality of the gravitational interaction.

\section{Connection and Tetrad Transformations}

A general element of the Lorentz group is written as 
\be
U_J = U_L \, U_S = U_S \, U_L \equiv \exp \left[- \ihalf \, \epsilon^{AB} \, J_{AB} \right],
\ee
with
\be %
U_S = \exp  \left[ - \ihalf \, \epsilon^{AB} \, S_{AB} \right] \quad \mbox{and} \quad
U_L = \exp  \left[ - \ihalf \, \epsilon^{AB} \, L_{AB} \right].
\label{u}
\ee %
By construction, under a local Lorentz transformation generated by $U_J$, the gauge covariant
derivative (\ref{locode}) transforms according to
\be
\Dbol'_{C'} \Psi'(x) = U_J \, \Dbol_C \Psi(x).
\label{transa}
\ee
Notice that, in addition to the Lorentz rotation in the matrix (spin) indices of $\Dbol_C$,
which is the only transformation occurring in (internal) Yang-Mills theories, in the case of
the (external) Lorentz gauge group the spacetime index of $\Dbol_C$ is also necessarily
transformed. Using the expressions \cite{ramond}
\[
\Psi(x) = U^{-1}_S \, \Psi'(x') \quad \mbox{and} \quad \Psi'(x) = U_L \,
\Psi'(x')
\]
in the transformation (\ref{transa}), it acquires the form
\be
U^{-1}_L \, \Dbol'_{C'} \, U_L = h_C{}^\mu \, U_S \, \Dbol_\mu \, U^{-1}_S.
\label{fitrans2}
\ee
Now, the Fock--Ivanenko derivative $\Dbol_\mu$ is defined by
\[
\Dbol{}_\mu \Psi = \partial_\mu \Psi +
\onehalf \, \Abol^{AB}{}_\mu \,
\frac{\delta_S \Psi}{\delta \epsilon^{AB}},
\]
where
\be
\delta_S \Psi \equiv \Psi^\prime(x^\prime) - \Psi(x) =
- \ihalf \, \epsilon^{AB} S_{AB} \Psi
\label{tlt1}
\ee
is the {\em total} change in $\Psi(x)$. It transforms, consequently, as
\be
\Dbol^\prime_\mu = U_S \, \Dbol_\mu \, U_S^{-1},
\label{figtrans}
\ee
from where we can obtain the typical connection gauge transformation
\be %
\Abol^\prime_\mu = U_S \Abol_\mu U_S^{-1} + i U_S \partial_\mu U_S^{-1}.
\label{trans1}
\ee %
Its infinitesimal version is given by
\be %
\delta_S \Abol^{CD}{}_\mu = - \left(\partial_\mu \epsilon^{CD} +
\Abol^C{}_{A \mu} \, \epsilon^{AD} + \Abol^D{}_{A \mu} \, \epsilon^{CA}
\right) \equiv - \Dbol_\mu \epsilon^{CD}.
\label{atrans}
\ee %
We see in this way that the Lorentz gauge transformations are generated by the spin (matrix)
part of the representation, that is, by the generators $S_{AB}$.

Let us then return to the transformation law (\ref{fitrans2}). Substituting
Eq.~(\ref{figtrans}), it becomes
\be
U^{-1}_L \, \Dbol'_{C'} \, U_L = h_C{}^\mu \, \Dbol'_\mu \equiv \Dbol'_C.
\label{fitrans3}
\ee
It is clear from this expression that the orbital part of the generators are responsible for
the Lorentz transformation in the spacetime index of the covariant derivative $\Dbol_{C}$. In
fact, denoting by $\Lambda^{A'}{}_C \equiv (U_S)^{A'}{}_C$ the element of the Lorentz group in
the vector representation, the tetrad transformation can be written in the usual form
\[
h_C{}^\mu = \Lambda^{A'}{}_C \; h_{A'}{}^\mu,
\]
and we easily see that
\be
\Dbol'_{C'} = \Lambda_{C'}{}^A \; \Dbol'_{A} \equiv U_L \, \Dbol'_{C} \, U^{-1}_L.
\label{fitrans4}
\ee
The transformation law of the covariant derivative can then be written as
\be
\Dbol^\prime_{C'} = \Lambda_{C'}{}^A \; U_S \, \Dbol_A \, U_S^{-1}.
\label{figtransbis}
\ee
It is important to observe the different roles played by each one of the Lorentz generators:
whereas the spin part $S_{AB}$ generates the (internal) Lorentz gauge transformation, the
orbital part $L_{AB}$ is responsible for transformation of the (external) spacetime index
of the covariant derivative.

Let us obtain now the infinitesimal Lorentz transformation of the tetrad field~(\ref{tetrada}).
The transformation generated by $S_{AB}$ corresponds to the {\em total} change in the tetrad,
that is, $\delta_S h^A{}_\mu \equiv h^{\prime A}{}_\mu(x^\prime) - h^A{}_\mu(x)$. From
Eq.~(\ref{tetrada}) we see that
\be %
\delta_S h^A{}_\mu = \partial_\mu (\delta_S x^A) + (\delta_S \Abol^A{}_{d \mu}) \,
x^D + \Abol^A{}_{D \mu} \, (\delta_S x^D).
\label{horbi}
\ee %
Using Eqs.~(\ref{slore}) and (\ref{atrans}), it is easy to see that
\be %
\delta_S h^A{}_\mu = \epsilon^A{}_C \, h^C{}_\mu \equiv - \ihalf
\, \epsilon^{CD} \, \left( S_{CD}\right)^A{}_B \, h^B{}_\mu,
\label{usualt}
\ee %
as it should be, since $h^A{}_\mu$ is a Lorentz vector field in the algebraic index. On the
other hand, the transformation generated by $L_{CD}$ corresponds to a change in the
coordinate only, that is, $\delta_L h^A{}_\mu \equiv h^A{}_\mu(x^\prime) - h^A{}_\mu(x)$.
Since $\Abol^{CD}{}_{\mu}$ responds only to the spin representation $S_{CD}$, we see from
Eq.~(\ref{tetrada}) that
\be %
\delta_L h^A{}_\mu = \partial_\mu (\delta_L x^A) + \Abol^A{}_{D \mu} \, (\delta_L x^D).
\label{horbi2}
\ee %
Using the transformation (\ref{relore}), we get
\be %
\delta_L h^A{}_\mu = - \Dbol_\mu \xi^A.
\ee
It is then easy to see that $\delta_L h^A{}_\mu$ induces on the metric tensor (\ref{gmn})
the transformation
\be
\delta_L g_{\mu \nu} = - \nabol_\mu \xi_\nu - \nabol_\nu \xi_\mu,
\ee
where $\xi_\nu = \xi_A \, h^A{}_\nu$, and $\nabol_\mu$ is the covariant derivative in the
spacetime connection (\ref{lcc}). As is well known, this equation represents the response of
$g_{\mu \nu}$ to spacetime diffeomorphisms $x'^\mu = x^\mu + \xi^\mu(x)$~\cite{landau}. In
other words, the Lorentz transformations generated by $L_{AB}$ are equivalent to a spacetime
general coordinate transformation. This means essentially that the ordinary formulation of
general relativity is included as a particular case of this more general approach, which
allows from the very beginning the inclusion of both integer and half-integer spin fields.
Furthermore, we see that the energy-momentum tensor of any matter field must follow, through
Noether's theorem~\cite{kopo}, from the invariance of the corresponding Lagrangian under a
Lorentz transformation generated by the orbital generators only~\cite{cp2}.

Concerning this last point, it is interesting to notice the following property. Usually,
the functional derivative of a matter field lagrangian ${\mathcal L}$ in relation to the
spin connection gives the spin tensor. However, due to the dependence of the tetrad on the
spin connection,
\be
{\mathcal J}^\mu{}_{AB} = \frac{1}{h} \, \frac{\delta {\mathcal L}}
{\delta \Abol^{AB}{}_\mu}
\ee
will represent now the {\em total} angular momentum, that is, spin plus orbital. In fact, as
the tetrad depends on the spin connection, the above expression can be rewritten in the form
\be
{\mathcal J}^\mu{}_{AB} =
\frac{1}{h} \, \frac{\delta {\mathcal L}}{\delta h^C{}_\rho} \;
\frac{\delta h^C{}_\rho} {\delta \Abol^{AB}{}_\mu}.
\label{amt2}
\ee
However, from Eq.~(\ref{tetrada}) we see that
\be
\frac{\delta h^C{}_\rho} {\delta \Abol^{AB}{}_\mu} =
\delta^\mu{}_\rho \, (\delta^C{}_A \, x_B - \delta^C{}_B \, x_A).
\ee
Substituting this relation in Eq.~(\ref{amt2}), we get
\be
{\mathcal J}^\mu{}_{AB} = x_A \, \Theta^\mu{}_B - x_B \,
\Theta^\mu{}_A,
\label{xamt}
\ee
where
\be
\Theta^\mu{}_A = - \frac{1}{h} \,
\frac{\delta {\mathcal L}}{\delta h^A{}_\mu}
\label{semt}
\ee
is the so called dynamical energy-momentum tensor. As far as ${\mathcal L}$ is local Lorentz
invariant, $\Theta^{\mu \nu} = \Theta^\mu{}_A \, h^{A \nu}$ will be the symmetric
energy--momentum tensor (see Ref.~\cite{weinberg}, page 371), and ${\mathcal J}^\mu{}_{AB}$
will represent the {\em total} angular momentum tensor~\cite{hay}.

\section{The Tangent-Space Coordinates as a Vector Field}

\subsection{General Properties}

As we have seen, the Minkowski coordinates $x^A$ can be interpreted both as a set of four
scalar functions---in which case its Lorentz transformation is that generated by the orbital
generators $L_{AB}$---and as a vector field $x^A(x^\mu)$, in which case its Lorentz
transformation is that generated by the vector representation $S_{AB}$. When we consider it as
a vector field, its Fock--Ivanenko covariant derivative
\be %
\Dbol_\mu x^C = \partial_\mu x^C +
\onehalf \, \Abol^{AB}{}_\mu \,
\frac{\delta_S x^C}{\delta \epsilon^{AB}} 
\ee %
turns out to be, through the use of the transformation (\ref{slore}), 
\be %
\Dbol_\mu x^C = \partial_\mu x^C +
\Abol^C{}_{B \mu} x^B \equiv h^C{}_\mu.
\label{dxh}
\ee %
This is to say that the Fock--Ivanenko covariant derivative of the vector field $x^C(x^\mu)$
exactly coincides with the tetrad field. We can consequently conclude, taking into account 
the usual concepts underlying the minimal coupling prescription, that the gravitational field
$\Abol^C{}_{D \mu}$ couples to the Minkowski tangent space coordinates, that is, to the vector
field $x^A$, and that the corresponding coupling constant is equal to 1. On their side, the
coordinates $\{x^{A}\}$ respond to the Lorentz gauge interaction. This is an essential point:
it is this coupling the responsible for the non--triviality of the tetrad, that is, for its
deviation from an exact differential form. A trivial, exact tetrad would represent a mere
coordinate transformation, and not a gravitational field. By the standard expression $g_{\mu
\nu} = \eta_{AB} \, h^A{}_\mu \, h^B{}_\nu$, therefore, it would give simply the Minkowski
metric written in arbitrary coordinates, with vanishing  Riemannian curvature. It is the
coordinate--connection coupling, with the ensuing tetrad non-triviality, that leads to the
non-trivial Riemannian metric on which the general relativity description of gravitation is
based. We can say, therefore, that the very foundation of that description is the coupling
between the spin connection and the tangent space coordinates. In addition, since the
nontrivial part $\Bbol^A{}_\mu = {\Abol^A}_{B \mu} \, x^B$  of the tetrad field comes from the
orbital part of the Lorentz generators (see Eq.~(\ref{nonholo})), and as its action takes
place on the {\em arguments} of every field, the {\em orbital} generators $L_{AB}$ appear as
the responsible for the universality of gravitation.

\subsection{Modified Palatini Action}

Using the fact that the tetrad is the covariant derivative of the tangent space coordinates,
as given by Eq.~(\ref{dxh}), the modified action functional of the gravitational field can be
rewritten in the form
\be
S'_{P} = \frac{1}{4 k^2} \int_{\mathcal R}
{\epsilon}^{\mu\nu\rho\sigma}\epsilon_{ABCD} \,
\Dbol_\mu x^A \, \Dbol_\nu x^B \, \Obol^{CD}{}_{\rho
\sigma}.
\label{palactionx}
\ee
Variation of $S'_{P}$ in relation to the spin connection $A^{AB}{}_\mu$ yields, up to a factor
of $x^A$, Einstein's equation. Now, using the tetrad (\ref{dxh}), as well as expression
(\ref{cur}) for the specific case of the Ricci coefficient of rotation, the constraint
equation (\ref{vincu}) acquires the form
\be
\Tbol^A{}_{\mu \nu} \equiv \Obol^A{}_{B \mu \nu} x^B = 0.
\label{rxt}
\ee
In a sense, torsion appears as a measure of the non-orthogonality between curvature and the
Minkowski coordinate field. When torsion vanishes, like in general relativity, curvature
becomes orthogonal to $x^A$.

On the other hand, as a consequence of the above described coupling between the spin
connection $\Abol^A{}_{B \mu}$ and the coordinates $x^A$, the gravitational action results to
depend also on the vector field $x^A$. It is then interesting to observe that a variation of
(\ref{palactionx}) in relation to $x^A$ yields
\be
\nabol_\mu\left[\Obol^\mu{}_\nu - \onehalf \, \delta^\mu{}_\nu
\Obol\right]=0,
\ee
which is the contracted form of the curvature Bianchi identity of general relativity.

\subsection{Lagrangian of the Tangent-Space Coordinates}

Now, if the coordinate $x^a(x^\mu)$ is considered to be a vector field, it must have a
Lagrangian density. Assuming that it is massless, the natural lagrangian for a vector field
on a Riemannian manifold, and responding to a gauge interaction, is
\be %
{\mathcal L}_x = - \epsilon_\Lambda \; \frac{h}{2} \ \eta_{CD} \, \eta^{AB}
\, \Dbol_A x^C \, \Dbol_B x^D,
\ee %
where $\epsilon_\Lambda$ is a positive constant with dimension of energy density, introduced
to give the Lagrangian the appropriate dimension. Using that
\be
\Dbol_A x^C = h^\mu{}_A \Dbol_\mu x^C \equiv h_A{}^\mu \, h^C{}_\mu =
\delta_A{}^C,
\ee
which follows from the orthogonality properties $h^A{}_\mu \, h_A{}^\nu = \delta_\mu{}^\nu$
and $h^A{}_\mu \, h_B{}^\mu = \delta^A{}_B$, we find
\be %
{\mathcal L}_x = - 2 \, h \, \epsilon_\Lambda.
\label{xla}
\ee %
The Lagrangian of the vector field $x^A$, therefore, corresponds to a cosmological term
for the gravitational field equations, with
\be
\Lambda = \frac{16 \pi G}{c^3} \, \epsilon_\Lambda
\ee
playing the role of cosmological constant. In other words, a cosmological term is nothing but a
Lagrangian for the tangent space coordinates, seen as a vector field. This result provides a
new interpretation for the origin of the dark energy of the universe~\cite{dark}, which would
then come from considering the tangent-space coordinates as a vector field. Observe that the
value of $\Lambda$ does not depend on the gravitational field, and is left as a free
parameter. In fact, even in the absence of gravitation, where the tetrads become trivial,
provided the tangent-space coordinates are considered as a vector field, its Lagrangian will
always give rise to a cosmological term.

\section{Final Remarks}

When general relativity is conceived as a genuine Lorentz-valued connection theory, the local
Lorentz group emerges as the \emph{kinematic} symmetry behind gravitation, and the Ricci
coefficient of rotation $\Abol{}^C{}_{D \mu}$, the spin connection of general relativity,
becomes the fundamental field describing gravitation. In this theory, a tetrad given by the
Fock-Ivanenko covariant derivative of the tangent space coordinates,
\be %
h^C{}_\mu \equiv \Dbol_\mu x^C = \partial_\mu x^C +
\Abol^C{}_{B \mu} x^B,
\label{tetra}
\ee %
shows up naturally. Since it depends on the spin connection, it is not a fundamental but a
derived field, as it should be in a true connection-based theory. A fundamental consequence of
this approach refers to the Bianchi identities. To see that, it is important to observe that
absence of torsion, as in Yang--Mills theories, is completely different from a present, but
vanishing torsion, as in general relativity. This difference is revealed by the fact that,
whereas in Yang--Mills theories there is only one Bianchi identity, in general relativity there
are two: one for torsion, given by
\be
\Obol^A{}_{\rho \mu \nu} +
\Obol^A{}_{\nu \rho \mu} + \Obol^A{}_{\mu \nu \rho} = 0,
\label{bi1}
\ee
and one for curvature, which reads
\be
\Dbol_\rho \Obol^A{}_{B \mu \nu} + \Dbol_\nu \Obol^A{}_{B
\rho \mu} + \Dbol_\mu \Obol^A{}_{B \nu \rho} = 0.
\label{bi2}
\ee
Multiplying the curvature Bianchi identity (\ref{bi2}) by $x^B$, and using the identity
(\ref{rxt}), the result is easily seen to be the torsion Bianchi identity (\ref{bi1}).
According to this approach, therefore, similarly to the Yang--Mills theories, general
relativity turns out to present only one independent Bianchi identity.

It is interesting to observe that, even in a general case, characterized by the simultaneous
presence of curvature and torsion, the two Bianchi identities are also reduced to a single one.
In this case, torsion satisfies the relation
\be
{\mathcal T}^A{}_{\mu \nu} \equiv \Omega^A{}_{B \mu \nu} \, x^B,
\label{rxtbis}
\ee
and is consequently always orthogonal to $x_A$:
\be
x_A \, {\mathcal T}^A{}_{\mu \nu} = 0.
\ee
Now, in this general case, the two Bianchi identities~\cite{w} can be written in the
form
\be
{\mathcal D}_\rho {\mathcal T}^A{}_{\mu \nu} +
{\mathcal D}_\nu {\mathcal T}^A{}_{\rho \mu} +
{\mathcal D}_\mu {\mathcal T}^A{}_{\nu \rho} =
\Omega^A{}_{\rho \mu \nu} +
\Omega^A{}_{\nu \rho \mu} + \Omega^A{}_{\mu \nu \rho}
\label{bibi1}
\ee
and
\be
{\mathcal D}_\rho \Omega^A{}_{B \mu \nu} + {\mathcal D}_\nu \Omega^A{}_{B
\rho \mu} + {\mathcal D}_\mu \Omega^A{}_{B \nu \rho} = 0.
\label{bibi2}
\ee
As before, multiplying the curvature Bianchi identity (\ref{bibi2}) by $x^B$, and using
the relation (\ref{rxtbis}), the result will be the torsion Bianchi identity (\ref{bibi1}).

Summing up, we can say that, when the fundamental field of gravitation is assumed to be the
spin connection, the two Bianchi identities of gravitation are reduced to a single one. This
result may be interpreted as an indication that, according to this approach, curvature and
torsion might be equivalent ways of describing the gravitational field, and consequently
related with the same degrees of freedom of gravity~\cite{torsion}. In this case, gravitation
becomes closer to the Yang--Mills theories which, as already pointed out, have only one
Bianchi identity. Whether the successful canonical quantization techniques of the Yang-Mills
theories will become applicable to such a theory is an open question yet to be explored.

\section*{Acknowledgments}
The authors would like to thank V. C. de Andrade for useful discussions. They would like also
to thank FAPESP-Brazil, CNPq-Brazil, CAPES-Brazil and COL\-CIENCIAS-Colombia for financial
support.

\end{document}